\documentclass{PoS}
\usepackage{graphicx}

\newcommand{\half}{{\frac{1}{2}}}
\newcommand{\mbf}[1]{\mathbf{#1}}
\renewcommand{\bar}[1]{\overline{#1}}

\title{QCD and Light-Front Dynamics}

\ShortTitle{QCD and Light-Front Dynamics}

\author{\speaker{Stanley J. Brodsky}
 \\
 SLAC National Accelerator Laboratory\\
Stanford University, Stanford, CA 94309, USA, and \\
CP$^3$-Origins,
Southern Denmark University, Odense, Denmark\\
 E-mail: \email{ sjbth@slac.stanford.edu}}

\author {Guy F. de T\'eramond\\
  Universidad de Costa Rica, San Jos\'e, Costa Rica\\
       E-mail: \email{gdt@asterix.crnet.cr}}
       
\abstract{AdS/QCD, the correspondence between theories in a dilaton-modified five-dimensional anti-de Sitter space and confining field theories in physical space-time, provides a remarkable semiclassical model for hadron physics.   Light-front holography allows hadronic amplitudes in the AdS fifth dimension to be mapped to frame-independent light-front wavefunctions of hadrons in physical space-time. The result is a single-variable light-front Schr\"odinger equation which determines the eigenspectrum and the light-front wavefunctions of hadrons for general spin and orbital angular momentum. The coordinate $z$ in AdS space is uniquely identified with  a Lorentz-invariant  coordinate $\zeta$ which measures the separation of the constituents within a hadron at equal light-front time and determines the off-shell dynamics of the bound state wavefunctions as a function of the invariant mass of the constituents.  The hadron eigenstates generally have components with different orbital angular momentum; e.g.,  the proton eigenstate in AdS/QCD with massless quarks has $L=0$ and $L=1$ light-front Fock components with equal probability.   Higher Fock states with extra quark-anti quark pairs also arise.  The soft-wall model also predicts the form of the non-perturbative effective coupling and its $\beta$-function. The AdS/QCD model can be systematically improved  by using its complete orthonormal solutions to diagonalize the full QCD light-front Hamiltonian or by applying the Lippmann-Schwinger method  to systematically include QCD interaction terms. Some novel features  of QCD are discussed, including the consequences of confinement for quark and gluon condensates. A method for computing the hadronization of quark and gluon jets at the amplitude level is outlined.}

\FullConference{Light Cone 2010: Relativistic Hadronic and Particle Physics\\
		 June 14-18, 2010, 
		 Valencia, Spain}

\begin{document}

\section{Introduction to Light-Front Hamiltonian QCD}

One of the most important theoretical tools in atomic physics is the
Schr\"odinger wavefunction, which describes the quantum-mechanical
structure of  an atomic system at the amplitude level. Light-front
wavefunctions (LFWFs) play a similar role in quantum chromodynamics, 
providing a fundamental description of the structure and
internal dynamics of hadrons in terms of their constituent quarks
and gluons. The LFWFs of bound states in QCD are
relativistic generalizations of the Schr\"odinger wavefunctions of
atomic physics, but they are determined at fixed light-cone time
 $\tau = x^0 + x^3$, the time marked by the
front of a light wave~\cite{Dirac:1949cp} instead of the ordinary instant time $t = x^0$.

Light-front quantization is the ideal framework to describe the
structure of hadrons in terms of their quark and gluon degrees of
freedom. The simple structure of the light-front (LF) vacuum allows an unambiguous
definition of the partonic content of a hadron in QCD and of hadronic light-front wavefunctions, 
which relate its quark
and gluon degrees of freedom to their asymptotic hadronic state. The constituent spin and orbital angular momentum properties of the hadrons are also encoded in the LFWFs. Unlike instant-time quantization, the Hamiltonian equation of motion on the light-front is frame-independent and has a structure similar to the eigenmode equations in AdS space.
This makes a direct connection of QCD with AdS/CFT methods possible. The identification of orbital angular momentum of the constituents is a key element in our description of the internal structure of hadrons using holographic principles, 
since hadrons with the same quark content, but different orbital angular momenta, have different masses.

A physical hadron in four-dimensional Minkowski space has four-momentum $P_\mu$ and invariant
hadronic mass states determined by the light-front
Lorentz-invariant Hamiltonian equation for the relativistic bound-state system
$P_\mu P^\mu \vert  \psi(P) \rangle = M^2 \vert  \psi(P) \rangle$, 
where the operator $P_\mu P^\mu$ is determined canonically from the QCD Lagrangian.

On AdS space the physical  states are
represented by normalizable modes $\Phi_P(x, z) = e^{-iP \cdot x} \Phi(z)$,
with plane waves along Minkowski coordinates $x^\mu$ and a profile function $\Phi(z)$ 
along the holographic coordinate $z$. The hadronic invariant mass 
$P_\mu P^\mu = M^2$  is found by solving the eigenvalue problem for the
AdS wave equation. Each  light-front hadronic state $\vert \psi(P) \rangle$ is dual to a normalizable string mode $\Phi_P(x,z)$. 
For fields near the AdS boundary the behavior of $\Phi(z)$
depends on the scaling dimension of corresponding interpolating operators.

We have recently shown a remarkable
connection between the description of hadronic modes in AdS space and
the Hamiltonian formulation of QCD in physical space-time quantized
on the light-front at equal light-front time  $\tau$.~\cite{deTeramond:2008ht}  In fact, one may take  the LF bound state Hamiltonian equation of motion in QCD as a starting point to  derive  relativistic wave equations in terms of an invariant transverse variable $\zeta$ which measures the
separation of the quark and gluonic constituents within the hadron
at the same LF time. The result is a single-variable light-front relativistic
Schr\"odinger equation,  which is
equivalent to the equations of motion which describe the propagation of spin-$J$ modes in a fixed  gravitational background asymptotic to AdS space.  Its eigenvalues give the hadronic spectrum and its eigenmodes represent the probability distribution of the hadronic constituents at a given scale.  Remarkably, the AdS equations correspond to the kinetic energy terms of  the partons inside a hadron, whereas the interaction terms build confinement and
correspond to the truncation of AdS space in an effective dual gravity  approximation.~\cite{deTeramond:2008ht}

Light-Front Holography is one of the most remarkable features of AdS/CFT.~\cite{Maldacena:1997re}  It  allows one to project the functional dependence of the wavefunction $\Phi(z)$ computed  in the  AdS fifth dimension to the  hadronic frame-independent light-front wavefunction $\psi(x_i, \mbf{b}_{\perp i})$ in $3+1$ physical space-time. The 
variable $z $ maps  to the LF variable $ \zeta(x_i, \mbf{b}_{\perp i})$. To prove this, we have shown that there exists a correspondence between the matrix elements of the electromagnetic current and the energy-momentum tensor of the fundamental hadronic constituents in QCD with the corresponding transition amplitudes describing the interaction of string modes in anti-de Sitter space with the external sources which propagate in the AdS interior. The agreement of the results for both 
electromagnetic~\cite{Brodsky:2006uqa, Brodsky:2007hb} and gravitational~\cite{Brodsky:2008pf} hadronic transition amplitudes provides an important consistency test and verification of holographic mapping from AdS to physical observables defined on the light-front.   The transverse coordinate $\zeta$ is closely related to the invariant mass squared  of the constituents in the LFWF  and its off-shellness  in  the LF kinetic energy,  and it is thus the natural variable to characterize the hadronic wavefunction.  In fact $\zeta$ is the only variable to appear in the relativistic light-front Schr\"odinger equations predicted from AdS/QCD in the limit of zero quark masses.

\section{Hadron Dynamics on the Light-Front}

A remarkable feature of LFWFs is the fact that they are frame
independent; i.e., the form of the LFWF is independent of the
hadron's total momentum $P^+ = P^0 + P^3$ and $\mbf{P}_\perp.$
The simplicity of Lorentz boosts of LFWFs contrasts dramatically with the complexity of the boost of wavefunctions defined at fixed time $t.$~\cite{Brodsky:1968ea}  
Light-front quantization is thus the ideal framework to describe the
structure of hadrons in terms of their quark and gluon degrees of freedom.  The
constituent spin and orbital angular momentum properties of the
hadrons are also encoded in the LFWFs.  
The total  angular momentum projection~\cite{Brodsky:2000ii} 
$J^z = \sum_{i=1}^n  S^z_i + \sum_{i=1}^{n-1} L^z_i$ 
is conserved Fock-state by Fock-state and by every interaction in the LF Hamiltonian.
Other advantageous features of light-front quantization include:

\begin{itemize}

\item
The simple structure of the light-front vacuum allows an unambiguous
definition of the partonic content of a hadron in QCD.  The chiral and gluonic condensates are properties of the higher Fock states,~\cite{Casher:1974xd,Brodsky:2009zd} rather than the vacuum.  In the case of the Higgs model, the effect of the usual Higgs vacuum expectation value is replaced by a constant $k^+=0$ zero mode field.~\cite{Srivastava:2002mw}

\item 
If one quantizes QCD in the physical light-cone gauge (LCG) $A^+ =0$, then gluons only have physical angular momentum projections $S^z= \pm 1$. The orbital angular momenta of quarks and gluons are defined unambiguously, and there are no ghosts.  

\item
The gauge-invariant distribution amplitude $\phi(x,Q)$  is the integral of the valence LFWF in LCG integrated over the internal transverse momentum $k^2_\perp < Q^2$, because the Wilson line is trivial in this gauge. It is also possible to quantize QCD in  Feynman gauge in the light front.~\cite{Srivastava:1999gi}

\item
LF Hamiltonian perturbation theory provides a simple method for deriving analytic forms for the analog of Parke-Taylor amplitudes~\cite{Motyka:2009gi} where each particle spin $S^z$ is quantized in the LF $z$ direction.  The gluonic $g^6$ amplitude  $T(-1 -1 \to +1 +1 +1 +1 +1 +1)$  requires $\Delta L^z =8;$ it thus must vanish at tree level since each three-gluon vertex has  $\Delta L^z = \pm 1.$ However, the order $g^8$ one-loop amplitude can be nonzero.

\item
Amplitudes in light-front perturbation theory are automatically renormalized using the ``alternate denominator"  subtraction method.~\cite{Brodsky:1973kb}  The application to QED has been checked at one and two loops.~\cite{Brodsky:1973kb}

\item 
One can easily show using LF quantization that the anomalous gravitomagnetic moment $B(0)$  of a nucleon, as  defined from the spin flip matrix element of the energy-momentum tensor, vanishes Fock-state by Fock state,~\cite{Brodsky:2000ii} as required by the equivalence principle.~\cite{Teryaev:1999su}

\item
LFWFs obey the cluster decomposition theorem, providing the only proof of this theorem for relativistic bound states.~\cite{Brodsky:1985gs}

\item
The LF Hamiltonian can be diagonalized using the discretized light-cone quantization (DLCQ) method.~\cite{Pauli:1985ps} This nonperturbative method is particularly useful for solving low-dimension quantum field theories such as 
QCD$(1+1).$~\cite{Hornbostel:1988fb}

\item 
LF quantization provides a distinction between static  (square of LFWFs) distributions versus non-universal dynamic structure functions,  such as the Sivers single-spin correlation and diffractive deep inelastic scattering which involve final state interactions.  The origin of nuclear shadowing and process independent anti-shadowing also becomes explicit.   This is discussed further in Sec.
\ref{rescat}.

\item 
LF quantization provides a simple method to implement jet hadronization at the amplitude level.  This is discussed in Sec. 
\ref{hadronization}.

\item 
The instantaneous fermion interaction in LF  quantization provides a simple derivation of the $J=0$
fixed pole contribution to deeply virtual Compton scattering.~\cite{Brodsky:2009bp}

\item
Unlike instant time quantization, the Hamiltonian equation of motion in the LF is frame independent. This makes a direct connection of QCD with AdS/CFT methods possible.~\cite{deTeramond:2008ht}

\end{itemize}

\section{Light-Front Holography}

A form factor in QCD is defined by the transition matrix element of a local quark current between hadronic states.  In AdS space form factors are computed from the overlap integral of normalizable modes with boundary currents which propagate in AdS space. The AdS/CFT duality incorporates the connection between the twist scaling dimension of the  QCD boundary interpolating operators
to the falloff of the of the normalizable modes in AdS near its conformal boundary. If both quantities represent the same physical observable for any value of the transfer momentum $q^2$, a precise correspondence can be established between the string modes $\Phi$ in AdS space and the light front wavefunctions of hadrons $\psi_{n/H}$ in physical four dimensional space-time.~\cite{Brodsky:2006uqa} The same results follow from comparing the relativistic light-front Hamiltonian equation describing bound states in QCD with the wave equations describing the propagation of modes in a warped AdS 
space.~\cite{deTeramond:2008ht}  In fact, one can systematically reduce  the LF  Hamiltonian equation to an effective relativistic wave equation, analogous to the AdS equations, by observing that each $n$-particle Fock state has an essential dependence on the invariant mass of the system  and
and thus, to a first approximation, LF dynamics depend only on the invariant mass of the system.
In  impact space the relevant variable is a boost-invariant  variable $\zeta$
which measures the separation of the constituents at equal LF time.

\subsection{Electromagnetic Form Factor \label{EMFF}}

Light-Front Holography can be derived by observing the correspondence between matrix elements obtained in AdS/CFT with the corresponding formula using the LF 
representation.~\cite{Brodsky:2006uqa}  The light-front electromagnetic form factor in impact 
space~\cite{Brodsky:2006uqa,Brodsky:2007hb,Soper:1976jc} can be written as a sum of overlap of light-front wave functions of the $j = 1,2, \cdots, n-1$ spectator
constituents:
\begin{equation} \label{eq:FFb}
F(q^2) =  \sum_n  \prod_{j=1}^{n-1}\int d x_j d^2 \mbf{b}_{\perp j}   \sum_q e_q  
            \exp \! {\Bigl(i \mbf{q}_\perp \! \cdot \sum_{j=1}^{n-1} x_j \mbf{b}_{\perp j}\Bigr)} 
 \left\vert  \psi_{n/H}(x_j, \mbf{b}_{\perp j})\right\vert^2 ,
\end{equation}
where the normalization is defined by
\begin{equation}  \label{eq:Normb}
\sum_n  \prod_{j=1}^{n-1} \int d x_j d^2 \mathbf{b}_{\perp j}
\vert \psi_{n/H}(x_j, \mathbf{b}_{\perp j})\vert^2 = 1.
\end{equation}

The formula  is exact if the sum is over all Fock states $n$.
For definiteness we shall consider a two-quark $\pi^+$  valence Fock state 
$\vert u \bar d\rangle$ with charges $e_u = \frac{2}{3}$ and $e_{\bar d} = \frac{1}{3}$.
For $n=2$, there are two terms which contribute to the $q$-sum in (\ref{eq:FFb}). 
Exchanging $x \leftrightarrow 1 \! - \! x$ in the second integral  we find 
\begin{equation}  \label{eq:PiFFb}
 F_{\pi^+}(q^2)  =  2 \pi \int_0^1 \! \frac{dx}{x(1-x)}  \int \zeta d \zeta \,
J_0 \! \left(\! \zeta q \sqrt{\frac{1-x}{x}}\right) 
\left\vert \psi_{u \bar d/ \pi}\!(x,\zeta)\right\vert^2,
\end{equation}
where $\zeta^2 =  x(1  -  x) \mathbf{b}_\perp^2$ and $F_{\pi^+}(q\!=\!0)=1$.

\begin{figure}[h]
\centering
\includegraphics[angle=0,width=10cm]{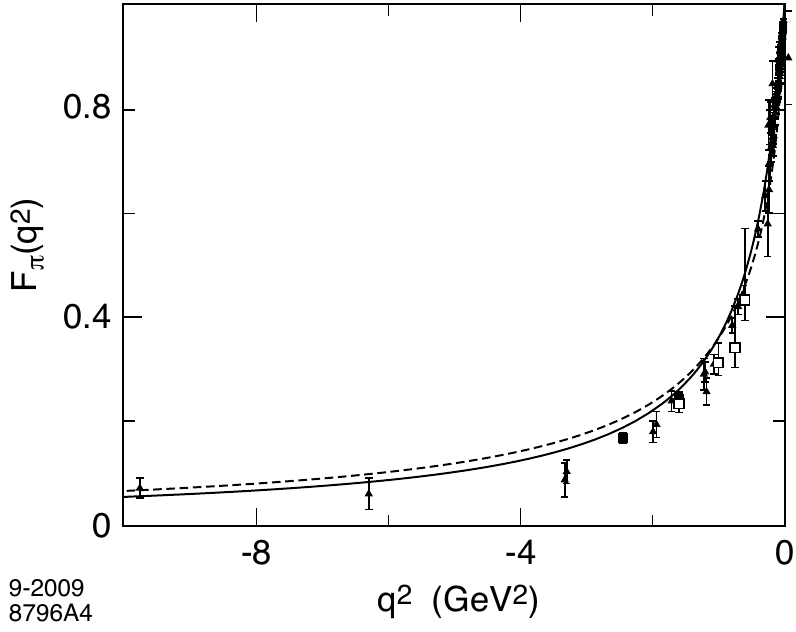}
\caption{Space-like scaling behavior for $F_\pi(Q^2)$ as a function of $q^2$.
The continuous line is the prediction of the soft-wall model for  $\kappa = 0.375$ GeV.
The dashed line is the prediction of the hard-wall model for $\Lambda_{\rm QCD} = 0.22$ GeV.
The triangles are the data compilation  from Baldini  {\it et al.},~\cite{Baldini:1998qn}  the  filled boxes  are JLAB 1 data~\cite{Tadevosyan:2007yd} and empty boxes are JLAB 2
 data.~\cite{Horn:2006tm}}
\label{PionFFSL}
\end{figure}

We now compare this result with the electromagnetic (EM) form-factor in AdS space:~\cite{Polchinski:2002jw}
\begin{equation} 
F(Q^2) = R^3 \int \frac{dz}{z^3} \, J(Q^2, z) \vert \Phi(z) \vert^2,
\label{eq:FFAdS}
\end{equation}
where $J(Q^2, z) = z Q K_1(z Q)$.
Using the integral representation of $J(Q^2,z)$
\begin{equation} \label{eq:intJ}
J(Q^2, z) = \int_0^1 \! dx \, J_0\negthinspace \left(\negthinspace\zeta Q
\sqrt{\frac{1-x}{x}}\right) ,
\end{equation} we write the AdS electromagnetic form-factor as
\begin{equation} 
F(Q^2)  =    R^3 \! \int_0^1 \! dx  \! \int \frac{dz}{z^3} \, 
J_0\!\left(\!z Q\sqrt{\frac{1-x}{x}}\right) \left \vert\Phi(z) \right\vert^2 .
\label{eq:AdSFx}
\end{equation}
Comparing with the light-front QCD  form factor (\ref{eq:PiFFb}) for arbitrary  values of $Q$~\cite{Brodsky:2006uqa}
\begin{equation} \label{eq:Phipsi} 
\vert \psi(x,\zeta)\vert^2 = 
\frac{R^3}{2 \pi} \, x(1-x)
\frac{\vert \Phi(\zeta)\vert^2}{\zeta^4}, 
\end{equation}
where we identify the transverse LF variable $\zeta$, $0 \leq \zeta \leq \Lambda_{\rm QCD}$,
with the holographic variable $z$.
AdS/QCD predictions for the space-like pion form factor are shown in Fig.  \ref{PionFFSL}.

Extension of the results to arbitrary $n$ follows from the $x$-weighted definition of the
transverse impact variable of the $n-1$ spectator system:~\cite{Brodsky:2006uqa}
\begin{equation} \label{zeta}
\zeta = \sqrt{\frac{x}{1-x}} ~ \Big\vert \sum_{j=1}^{n-1} x_j \mbf{b}_{\perp j} \Big\vert ,
\end{equation}
where $x = x_n$ is the longitudinal
momentum fraction of the active quark.
A recent application of the light-front holographic ideas has been used to compute the helicity-independent generalized parton distributions (GPDs) of quarks
in a  nucleon in the zero skewness case.~\cite{Vega:2010ns}

Conserved currents are not renormalized and correspond to five dimensional massless fields propagating in AdS according to the relation 
$(\mu R)^2 = (\Delta - p) (\Delta + p -  4)$  for a $p$ form in $d=4$. In the usual AdS/QCD framework~\cite{Erlich:2005qh, DaRold:2005zs} this  corresponds to $\Delta = 3$ or 1, the canonical dimensions of
an EM current and field strength respectively.  Normally one uses a hadronic  interpolating operator  with minimum twist $\tau$ to identify a hadron in AdS/QCD and to predict the power-law fall-off behavior of its form factors and other hard 
scattering amplitudes;~\cite{Polchinski:2001tt} {\it e.g.},  for a two-parton bound state $\tau = 2$.   However, in the case of a current, one needs to  use  an effective field operator  with dimension $\Delta =3.$ The apparent inconsistency between twist and dimension is removed by noticing that in the light-front one chooses to calculate the  matrix element of the twist-3 plus  component of the current  $J^+$,~\cite{Brodsky:2006uqa, Brodsky:2007hb} in order to avoid coupling to Fock states with different numbers of constituents.

\subsection{Gravitational Form Factor}

Matrix elements of the energy-momentum tensor $\Theta^{\mu \nu} $ which define the gravitational form factors play an important role in hadron physics.  Since one can define $\Theta^{\mu \nu}$ for each parton, one can identify the momentum fraction and  contribution to the orbital angular momentum of each quark flavor and gluon of a hadron. For example, the spin-flip form factor $B(q^2)$, which is the analog of the Pauli form factor $F_2(Q^2)$ of a nucleon, provides a  measure of the orbital angular momentum carried by each quark and gluon constituent of a hadron at $q^2=0.$   Similarly,  the spin-conserving form factor $A(q^2)$, the analog of the Dirac form factor $F_1(q^2)$, allows one to measure the momentum  fractions carried by each constituent.
This is the underlying physics of Ji's sum rule:~\cite{Ji:1996ek}
$\langle J^z\rangle = \half [ A(0) + B(0)] $,  which has prompted much of the current interest in 
the generalized parton distributions (GPDs)  measured in deeply
virtual Compton scattering.  An important constraint is $B(0) = \sum_i B_i(0) = 0$;  {\it i.e.}  the anomalous gravitomagnetic moment of a hadron vanishes when summed over all the constituents $i$. This was originally derived from the equivalence principle of gravity.~\cite{Teryaev:1999su}  The explicit verification of these relations, Fock state by Fock state, can be obtained in the LF quantization of QCD in  light-cone 
gauge.~\cite{Brodsky:2000ii} Physically $B(0) =0$ corresponds to the fact that the sum of the $n$ orbital angular momenta $L$ in an $n$-parton Fock state must vanish since there are only $n-1$ independent orbital angular momenta.

The LF expression for the helicity-conserving gravitational form factor in impact space
is~\cite{Brodsky:2008pf}
\begin{equation} \label{eq:Ab}
A(q^2) =  \sum_n  \prod_{j=1}^{n-1}\int d x_j d^2 \mbf{b}_{\perp j}  \sum_f  x_f 
\exp \! {\Bigl(i \mbf{q}_\perp \! \cdot \sum_{j=1}^{n-1} x_j \mbf{b}_{\perp j}\Bigr)} 
\left\vert  \psi_{n/H}(x_j, \mbf{b}_{\perp j})\right\vert^2,
\end{equation}
which includes the contribution of each struck parton with longitudinal momentum $x_f$
and corresponds to a change of transverse momentum $x_j \mbf{q}$ for
each of the $j = 1, 2, \cdots, n-1$ spectators. 
For $n=2$, there are two terms which contribute to the $f$-sum in  (\ref{eq:Ab}). 
Exchanging $x \leftrightarrow 1-x$ in the second integral we find 
\begin{equation} \label{eq:PiGFFb}
A_{\pi}(q^2) =  4 \pi \int_0^1 \frac{dx}{(1-x)}  \int \zeta d \zeta \,
J_0 \! \left(\! \zeta q \sqrt{\frac{1-x}{x}}\right) 
\left\vert \psi_{q \bar q/ \pi}\!(x,\zeta)\right\vert^2,
\end{equation}
where $\zeta^2 =  x(1-x) \mathbf{b}_\perp^2$ and  $A_{\pi}(0) = 1$.

 We now consider the expression for the hadronic gravitational form factor in AdS space, which is obtained by perturbing the metric from the static
 AdS geometry~\cite{Abidin:2008ku}
\begin{equation} 
A_\pi(Q^2)  =  R^3 \! \! \int \frac{dz}{z^3} \, H(Q^2, z) \left\vert\Phi_\pi(z) \right\vert^2,
\end{equation}
where $H(Q^2, z) = \half  Q^2 z^2  K_2(z Q)$ and $A(0) = 1$.
Using the integral representation of $H(Q^2,z)$
\begin{equation} \label{eq:intHz}
H(Q^2, z) =  2  \int_0^1\!  x \, dx \, J_0\!\left(\!z Q\sqrt{\frac{1-x}{x}}\right) ,
\end{equation}
we can write the AdS gravitational form factor 
\begin{equation} 
A(Q^2)  =  2  R^3 \! \int_0^1 \! x \, dx  \! \int \frac{dz}{z^3}  \,
J_0\!\left(\!z Q\sqrt{\frac{1-x}{x}}\right) \left \vert\Phi(z) \right\vert^2 .
\label{eq:AdSAx}
\end{equation}
Comparing with the QCD  gravitational form factor (\ref{eq:PiGFFb}) we find an  identical  relation between the LF wave function $\psi(x,\zeta)$ and the AdS wavefunction $\Phi(z)$
given in Eq. (\ref{eq:Phipsi}) which was obtained in Sect. \ref{EMFF} from the mapping of the pion electromagnetic transition amplitude.

As for the case of the electromagnetic form factor, the AdS mapping of the gravitational form factor is carried out in light-front holography for the plus-plus components of the energy-momentum tensor $\Theta^{++}$. The twist of this operator is $\tau = 4$ and coincides with the canonical conformal dimension of the energy-momentum tensor.

\subsection{Light-Front Bound-State Hamiltonian Equation of Motion}

A key step in the analysis of an atomic system such as positronium
is the introduction of the spherical coordinates $r, \theta, \phi$
which  separates the dynamics of Coulomb binding from the
kinematical effects of the quantized orbital angular momentum $L$.
The essential dynamics of the atom is specified by the radial
Schr\"odinger equation whose eigensolutions $\psi_{n,L}(r)$
determine the bound-state wavefunction and eigenspectrum. In our recent 
work, we have shown that there is an analogous invariant
light-front coordinate $\zeta$ which allows one to separate the
essential dynamics of quark and gluon binding from the kinematical
physics of constituent spin and internal orbital angular momentum.
The result is a single-variable LF Schr\"odinger equation for QCD
which determines the eigenspectrum and the light-front wavefunctions
of hadrons for general spin and orbital angular momentum.~\cite{deTeramond:2008ht}  If one further chooses  the constituent rest frame (CRF)~\cite{Danielewicz:1978mk,Karmanov:1979if,Glazek:1983ba}  where $\sum^n_{i=1} \mbf{k}_i \! = \! 0$, then the kinetic energy in the LFWF displays the usual 3-dimensional rotational invariance. Note that if the binding energy is nonzero, $P^z \ne 0,$ in this frame.

One can also derive light-front holography using a first semiclassical approximation  to transform the fixed 
light-front time bound-state Hamiltonian equation of motion in QCD
\begin{equation} \label{LFH}
H_{LF} \vert  \psi(P) \rangle =  \mathcal{M}_{H}^2 \vert  \psi(P) \rangle,
\end{equation}
with  $H_{LF} \equiv P_\mu P^\mu  =  P^- P^+ -  \mbf{P}_\perp^2$,
to  a corresponding wave equation in AdS 
space.~\cite{deTeramond:2008ht} To this end we
 compute the invariant hadronic mass $\mathcal{M}^2$ from the hadronic matrix element
\begin{equation}
\langle \psi_H(P') \vert H_{LF} \vert \psi_H(P) \rangle  = 
\mathcal{M}_H^2  \langle \psi_H(P' ) \vert\psi_H(P) \rangle,
\end{equation}
expanding the initial and final hadronic states in terms of its Fock components. We use the 
frame $P = \big(P^+, M^2/P^+, \vec{0}_\perp \big)$ where $H_{LF} =  P^+ P^-$.
We find 
\begin{equation}   
 \mathcal{M}_H^2  =  \sum_n  \prod_{j=1}^{n-1} \int d x_j \, d^2 \mbf{b}_{\perp j} \,
\psi_{n/H}^*(x_j, \mbf{b}_{\perp j}) 
  \sum_q   \left(\frac{ \mbf{- \nabla}_{ \mbf{b}_{\perp q}}^2  \! + m_q^2 }{x_q} \right) 
 \psi_{n/H}(x_j, \mbf{b}_{\perp j}) 
  + {\rm (interactions)} , \label{eq:Mba}
 \end{equation}
plus similar terms for antiquarks and gluons ($m_g = 0)$.

 Each constituent of the light-wavefunction  $\psi_{n/H}(x_i, \mbf{k}_{\perp i}, \lambda_i)$  of a hadron is on its respective mass shell 
 $k^2_i= k^+_i k^-_i - \mbf{k}^2_\perp = m^2_i$, $i = 1, 2 \cdots n.$   Thus $k^-= 
{{\mbf k}^2_\perp+  m^2_i\over x_i P^+}$.
However,  the light-front wavefunction represents a state which is off the light-front energy shell: $P^-  - \sum_i^n k^-_n < 0$, for a stable hadron.  Scaling out $P^+ = \sum^n_i k^+_i$, the 
off-shellness of the $n$-parton LFWF is thus $\mathcal{M}^2_H -  \mathcal{M}^2_n$, where
the invariant mass  of the constituents $\mathcal{M}_n$ is
\begin{equation}
 \mathcal{M}_n^2  = \Big( \sum_{i=1}^n k_i^\mu\Big)^2 = \sum_i \frac{\mbf{k}_{\perp i}^2 +  m_i^2}{x_i} ,
 \end{equation}

The action principle selects the configuration which minimizes the time-integral of the Lagrangian $L= T-V$, thus minimizing the kinetic energy $T$  and maximizing the attractive forces of the potential $V$. Thus in a fixed potential, the light-front wavefunction  peaks at the minimum value of the invariant mass of the constituents; i.e. at  the minimum off-shellness $M^2_H -  {\cal M}^2_n$.   In the case of massive constituents,  the minimum LF off-shellness occurs when all of the constituents have equal rapidity: $x_i \simeq {m^2_{\perp i}\over \sum^n_j m^2_{\perp j}}, $ where $m_{\perp i} = \sqrt {k^2_{\perp i} + m^2_i}.$ This is the central principle underlying the intrinsic heavy sea-quark distributions of hadrons.  The functional dependence  for a given Fock state is given in terms of the invariant mass,  the measure of the off-energy shell of the bound state.  

If we want to simplify further the description of the multiple parton system and reduce its dynamics to a single variable problem, we must take the limit of quark masses to zero. 
Indeed, the underlying classical QCD Lagrangian with massless quarks is scale and conformal invariant~\cite{Parisi:1972zy}, and consequently only in this limit it is possible to map the equations of motion and transition matrix elements to their correspondent conformal AdS  expressions.

 To simplify the discussion we will consider a two-parton hadronic bound state.  In the limit of zero quark masses
$m_q \to 0$
\begin{equation}  \label{eq:Mb}
\mathcal{M}^2  =  \int_0^1 \! \frac{d x}{x(1-x)} \int  \! d^2 \mbf{b}_\perp  \,
  \psi^*(x, \mbf{b}_\perp) 
  \left( - \mbf{\nabla}_{ {\mbf{b}}_{\perp}}^2\right)
  \psi(x, \mbf{b}_\perp) +   {\rm (interactions)}.
 \end{equation}
For $n=2$, ${\mathcal M}_{n=2}^2 = \frac{\mbf{k}_\perp^2}{x(1-x)}$. 
Similarly in impact space the relevant variable for a two-parton state is  $\zeta^2= x(1-x)\mbf{b}_\perp^2$.
Thus, to first approximation  LF dynamics  depend only on the boost invariant variable
$\mathcal{M}_n$ or $\zeta,$
and hadronic properties are encoded in the hadronic mode $\phi(\zeta)$ from the relation
\begin{equation} \label{eq:psiphi}
\psi(x,\zeta, \varphi) = e^{i M \varphi} X(x) \frac{\phi(\zeta)}{\sqrt{2 \pi \zeta}} ,
\end{equation}
thus factoring out the angular dependence $\varphi$ and the longitudinal, $X(x)$, and transverse mode $\phi(\zeta)$
with normalization $ \langle\phi\vert\phi\rangle = \int \! d \zeta \,
 \vert \langle \zeta \vert \phi\rangle\vert^2 = 1$. The mapping  of transition matrix elements
 for arbitrary values of the momentum transfer~\cite{Brodsky:2006uqa,Brodsky:2007hb,Brodsky:2008pf} 
 gives $X(x) = \sqrt{x(1-x)}$.

We can write the Laplacian operator in (\ref{eq:Mb}) in circular cylindrical coordinates $(\zeta, \varphi)$
and factor out the angular dependence of the
modes in terms of the $SO(2)$ Casimir representation $L^2$ of orbital angular momentum in the
transverse plane. Using  (\ref{eq:psiphi}) we find~\cite{deTeramond:2008ht}
\begin{equation} \label{eq:KV}
\mathcal{M}^2   =  \int \! d\zeta \, \phi^*(\zeta) \sqrt{\zeta}
\left( -\frac{d^2}{d\zeta^2} -\frac{1}{\zeta} \frac{d}{d\zeta}
+ \frac{L^2}{\zeta^2}\right)
\frac{\phi(\zeta)}{\sqrt{\zeta}}   \\
+ \int \! d\zeta \, \phi^*(\zeta) \, U(\zeta)  \, \phi(\zeta) ,
\end{equation}
where all the complexity of the interaction terms in the QCD Lagrangian is summed up in the effective potential $U(\zeta)$.
The light-front eigenvalue equation $H_{LF} \vert \phi \rangle = \mathcal{M}^2 \vert \phi \rangle$
is thus a LF wave equation for $\phi$
\begin{equation} \label{LFWE}
\left(-\frac{d^2}{d\zeta^2}
- \frac{1 - 4L^2}{4\zeta^2} + U(\zeta) \right) 
\phi(\zeta) = \mathcal{M}^2 \phi(\zeta),
\end{equation}
an effective single-variable light-front Schr\"odinger equation which is
relativistic, covariant and analytically tractable. Using (\ref{eq:Mb}) one can readily
generalize the equations to allow for the kinetic energy of massive
quarks.~\cite{Brodsky:2008pg}  In this case, however,
the longitudinal mode $X(x)$ does not decouple from the effective LF bound-state equations.

We now compare (\ref{LFWE}) with the wave equation in AdS$_{d+1}$ space for a spin-$J$ mode $\Phi_J$, $\Phi_J= \Phi_{\mu_1 \mu_2 \cdots \mu_J}$, with all the polarization indices
along the physical 3 + 1 coordinates~\cite{deTeramond:2008ht, deTeramond:2010ge}
\begin{equation} \label{WeJ}
\left[-\frac{ z^{d-1 -2 J}}{e^{\varphi(z)}}   \partial_z \left(\frac{e^{\varphi(z)}}{z^{d-1 - 2 J}} \partial_z\right)
+ \left(\frac{\mu R}{z}\right)^2\right] \Phi_{\mu_1 \mu_2 \cdots \mu_J} = M^2 \Phi_{\mu_1 \mu_2 \cdots \mu_J},
\end{equation}

Upon the substitution $z \! \to\! \zeta$  and
$\phi_J(\zeta)   = \left(\zeta/R\right)^{-3/2 + J} e^{\varphi(z)/2} \, \Phi_J(\zeta)$,
in (\ref{WeJ}), we find for $d=4$ the QCD light-front wave equation (\ref{LFWE}) with effective potential~\cite{deTeramond:2010ge}
\begin{equation} \label{U}
U(\zeta) = \half \varphi''(z) +\frac{1}{4} \varphi'(z)^2  + \frac{2J - 3}{2 z} \varphi'(z) ,
\end{equation}
where the fifth dimensional mass $\mu$ is not a free parameter but scales as $(\mu R)^2 = - (2-J)^2 + L^2$. If $L^2 \ge 0$ the LF Hamiltonian is positive definite
 $\langle \phi \vert H_{LF} \vert \phi \rangle \ge 0$ and thus $\mathcal M^2 \ge 0$.
 If $L^2 < 0$ the bound state equation is unbounded from below. The critical value corresponds to $L=0$.
 The quantum mechanical stability $L^2 >0$ for $J=0$ is thus equivalent to the
 Breitenlohner-Freedman stability bound in AdS.~\cite{Breitenlohner:1982jf}
 The AdS equations
correspond to the kinetic energy terms of  the partons inside a
hadron, whereas the interaction terms build confinement. 

In the hard-wall model one has $U(z)=0$; confinement is introduced by requiring the wavefunction to vanish at $z=z_0 \equiv 1/\Lambda_{\rm QCD}.$~\cite{Polchinski:2001tt}
In the case of the soft-wall model,~\cite{Karch:2006pv}  the potential arises from a  ``dilaton'' modification of the AdS metric; it  has the form of a harmonic oscillator. For the confining  positive-sign dilaton background $\exp(+ \kappa^2 z^2)$~\cite {deTeramond:2009xk, Andreev:2006ct}  we find the effective potential 
$U(z) = \kappa^4 z^2 + 2 \kappa^2(L+S-1)$. The resulting mass spectra  for mesons  at zero quark mass is
${\cal M}^2 = 4 \kappa^2 (n + L +S/2)$.

\begin{figure}[h]
\begin{center}
\includegraphics[width=7.2cm]{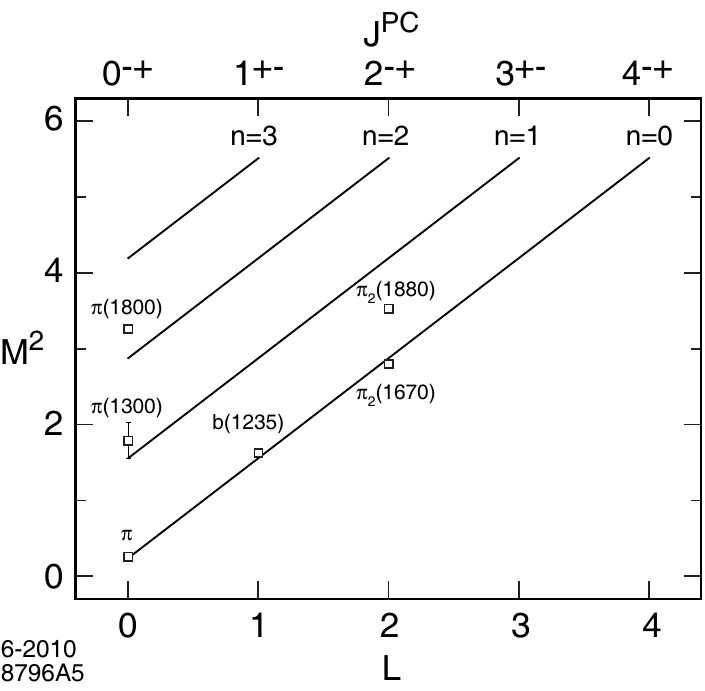}  \hspace{10pt}
\includegraphics[width=7.2cm]{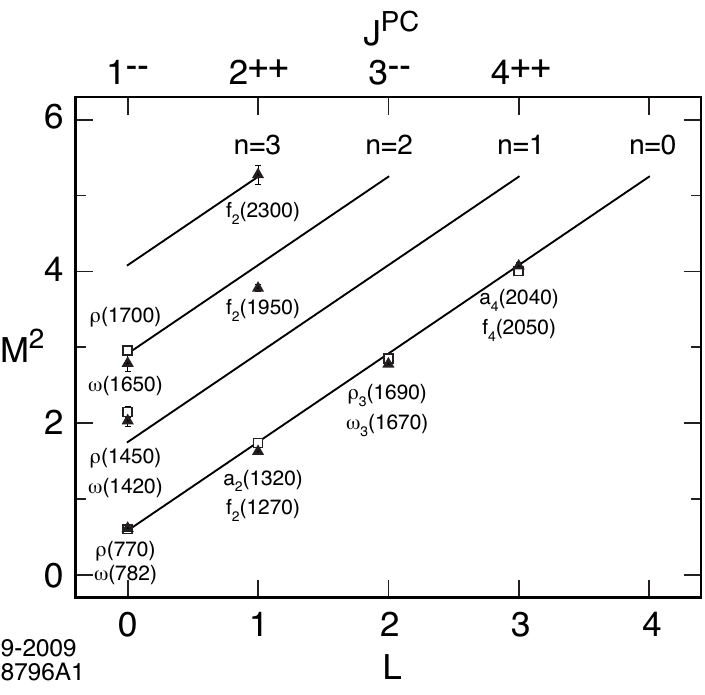}
 \caption{Parent and daughter Regge trajectories for (a) the $\pi$-meson family with
$\kappa= 0.6$ GeV; and (b) the  $I\!=\!1$ $\rho$-meson
 and $I\!=\!0$  $\omega$-meson families with $\kappa= 0.54$ GeV. Only confirmed PDG states~\cite{Amsler:2008xx} are shown.}
\label{pionspec}
\end{center}
\end{figure}

\begin{figure}[h]
\begin{center}
\includegraphics[width=10cm]{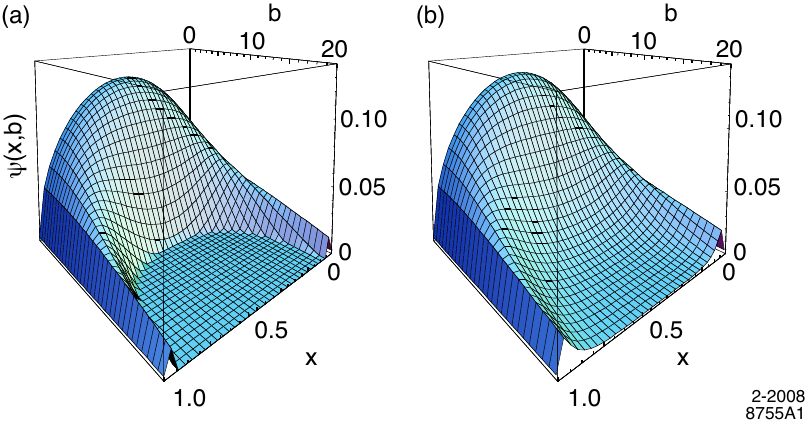}
 \caption{Pion light-front wavefunction $\psi_\pi(x, \mbf{b}_\perp$) for the  AdS/QCD (a) hard-wall and (b) soft-wall  models.}
\label{LFWF} 
\end{center} 
\end{figure}

The spectral predictions for  light meson and vector meson  states are compared with experimental data
in Fig. \ref{pionspec} for the positive sign dilaton model discussed here.
The corresponding wavefunctions for the hard and soft-wall models (see Fig.  \ref{LFWF}) 
display confinement at large interquark
separation and conformal symmetry at short distances, reproducing dimensional counting rules for hard exclusive amplitudes.

The predictions for the $\bf 56$-plet of light baryons under the $SU(6)$  flavor group are shown in Fig. \ref{baryons}. As for the predictions for mesons in Fig. \ref{pionspec}, only confirmed PDG~\cite{Amsler:2008xx} states are shown. 
The Roper state $N(1440)$ and the $N(1710)$ are well accounted for in this model as the first  and second radial
states. Likewise the $\Delta(1660)$ corresponds to the first radial state of the $\Delta$ family. The model is  successful in explaining the important parity degeneracy observed in the light baryon spectrum, such as the $L\! =\!2$, $N(1680)\!-\!N(1720)$ pair and the $\Delta(1905), \Delta(1910), \Delta(1920), \Delta(1950)$ states which are degenerate 
within error bars. The parity degeneracy of baryons is also a property of the hard wall model, but radial states are not well described by this model.~\cite{deTeramond:2005su}

\begin{figure}[h]
\begin{centering}
\includegraphics[angle=0,width=14.8cm]{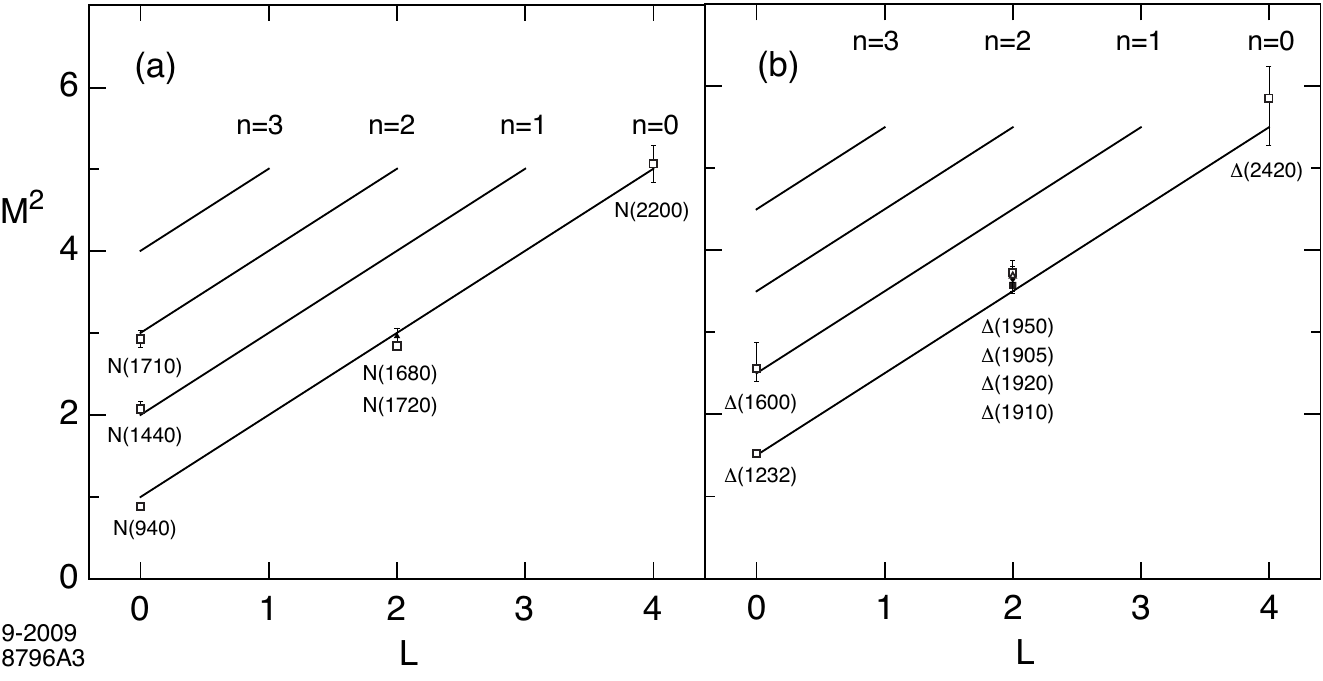}
\caption{\label{baryons}{{\bf 56} \small Parent and daughter Regge trajectories for  the  $N$ and $\Delta$ 
baryon families for $\kappa= 0.5$ GeV.
Data from \cite{Amsler:2008xx}. }
}
\label{baryons}
\end{centering}
\end{figure}
 
 An important feature of light-front holography is that it predicts the same multiplicity of states for mesons
and baryons as it is observed experimentally.~\cite{Klempt:2007cp} This remarkable property could have a simple explanation in the cluster decomposition of the
holographic variable $\zeta$ (\ref{zeta}), which labels a system of partons as an active quark plus a system on $n-1$ spectators. From this perspective a baryon with $n=3$ looks in light-front holography as a quark-diquark system.

Nonzero quark masses are naturally incorporated into the AdS/LF predictions~\cite{Brodsky:2008pg} by including them explicitly in the LF kinetic energy  $\sum_i ( {\mbf{k}^2_{\perp i} + m_i^2})/{x_i}$. Given the nonpertubative LFWFs one can predict many interesting phenomenological quantities such as heavy quark decays, generalized parton distributions and parton structure functions.  The AdS/QCD model is semiclassical, and thus it only predicts the lowest valence Fock state structure of the hadron LFWF. One can systematically improve the holographic approximation by
diagonalizing the QCD LF Hamiltonian on the AdS/QCD basis~\cite{Vary:2009gt}, or by using the Lippmann-Schwinger equations.
The action of the non-diagonal terms
in the QCD interaction Hamiltonian also generates the form of the higher
Fock state structure of hadronic LFWFs.  In
contrast with the original AdS/CFT correspondence, the large $N_C$
limit is not required to connect light-front QCD to
an effective dual gravity approximation.

\section {Vacuum Effects and Light-Front Quantization}

The LF vacuum is remarkably simple in light-front quantization because of the restriction $k^+ \ge 0.$   For example in QED,  vacuum graphs such as $e^+ e^- \gamma $  associated with the zero-point energy do not arise. In the Higgs theory, the usual Higgs vacuum expectation value is replaced with a $k^+=0$ zero mode;~\cite{Srivastava:2002mw} however, the resulting phenomenology is identical to the standard analysis.

Hadronic condensates play an important role in quantum chromodynamics (QCD).
Conventionally, these condensates are considered to be properties
of the QCD vacuum and hence to be constant throughout space-time.
Recently a new perspective on the nature of QCD
condensates $\langle \bar q q \rangle$ and $\langle
G_{\mu\nu}G^{\mu\nu}\rangle$, particularly where they have spatial and temporal
support,
has been presented.~\cite{Brodsky:2009zd,Brodsky:2008be,Brodsky:2008xm,Brodsky:2008xu} 
Their spatial support is restricted to the interior
of hadrons, since these condensates arise due to the interactions of quarks and
gluons which are confined within hadrons. For example, consider a meson consisting of a light quark $q$ bound to a heavy
antiquark, such as a $B$ meson.  One can analyze the propagation of the light
$q$ in the background field of the heavy $\bar b$ quark.  Solving the
Dyson-Schwinger equation for the light quark one obtains a nonzero dynamical
mass and, via the connection mentioned above, hence a nonzero value of the
condensate $\langle \bar q q \rangle$.  But this is not a true vacuum
expectation value; instead, it is the matrix element of the operator $\bar q q$
in the background field of the $\bar b$ quark.  The change in the (dynamical)
mass of the light quark in this bound state is somewhat reminiscent of the
energy shift of an electron in the Lamb shift, in that both are consequences of
the fermion being in a bound state rather than propagating freely.
Similarly, it is important to use the equations of motion for confined quarks
and gluon fields when analyzing current correlators in QCD, not free
propagators, as has often been done in traditional analyses of operator
products.  Since after a $q \bar q$ pair is created, the distance between the
quark and antiquark cannot get arbitrarily great, one cannot create a quark
condensate which has uniform extent throughout the universe. 
As a result, it is argued in Refs. ~\cite{Brodsky:2009zd,Brodsky:2008be,Brodsky:2008xm,Brodsky:2008xu}   that the 45 orders of magnitude conflict of QCD with the observed value of the cosmological condensate is removed
A new perspective on the nature of quark and gluon condensates in
quantum chromodynamics is thus obtained:~\cite{Brodsky:2008be,Brodsky:2008xm,Brodsky:2008xu}  the spatial support of QCD condensates
is restricted to the interior of hadrons, since they arise due to the
interactions of confined quarks and gluons.  In the LF theory, the condensate physics is replaced by the dynamics of higher non-valence Fock states as shown by Casher and Susskind.~\cite{Casher:1974xd}  In particular, chiral symmetry is broken in a limited domain of size $1/ m_\pi$,  in analogy to the limited physical extent of superconductor phases.  This novel description  of chiral symmetry breaking  in terms of ``in-hadron condensates"  has also been observed in Bethe-Salpeter studies.~\cite{Maris:1997hd,Maris:1997tm}
This picture explains the
results of recent studies~\cite{Ioffe:2002be,Davier:2007ym,Davier:2008sk} which find no significant signal for the vacuum gluon
condensate.

AdS/QCD also provides  a description of chiral symmetry breaking by
using the propagation of a scalar field $X(z)$
to represent the dynamical running quark mass. 
In the hard wall model the solution has the form~\cite{Erlich:2005qh,DaRold:2005zs} $X(z) = a_1 z+ a_2 z^3$, where $a_1$ is
proportional to the current-quark mass. The coefficient $a_2$ scales as
$\Lambda^3_{QCD}$ and is the analog of $\langle \bar q q \rangle$; however,
since the quark is a color nonsinglet, the propagation of $X(z),$ and thus the
domain of the quark condensate, is limited to the region of color confinement. 
Furthermore the effect of the $a_2$ term
varies within the hadron, as characteristic of an in-hadron condensate. 
The AdS/QCD picture of condensates with spatial support restricted to hadrons
is also in general agreement with results from chiral bag 
models,~\cite{Chodos:1975ix,Brown:1979ui,Hosaka:1996ee}
which modify the original MIT bag by coupling a pion field to the surface of
the bag in a chirally invariant manner.

\section{Hadronization at the Amplitude Level \label{hadronization}}

\begin{figure}[!]
\centering
\includegraphics[width=10cm]{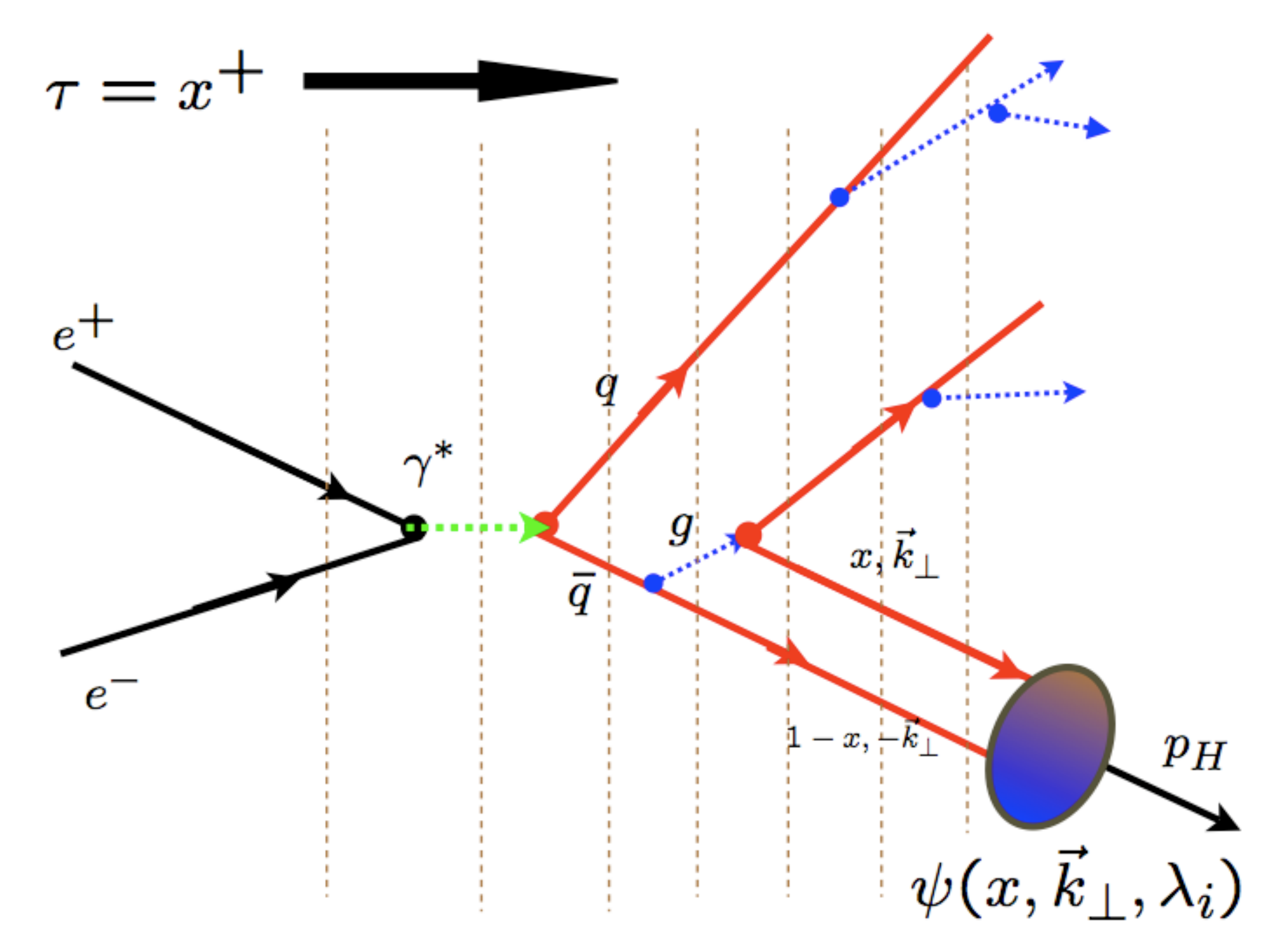}
  \caption{Illustration of an event amplitude generator for $e^+ e^- \to \gamma^* \to X$ for 
  hadronization processes at the amplitude level. Capture occurs if the quarks try to go beyond the confinement distance;  i.e, if
  $\zeta^2 = x(1-x) \mbf{b}_\perp^2 > 1/ \Lambda_{\rm QCD}^2$
   in the AdS/QCD hard-wall model of confinement. The corresponding condition in momentum space is
  $\mathcal{M}^2 = \frac{\mbf{k}_\perp^2}{x(1-x)} \lesssim \Lambda_{\rm QCD}^2$.}
\label{hadroniz}  
\end{figure}

The conversion of quark and gluon partons  to hadrons is usually discussed in terms  of on-shell hard-scattering cross sections convoluted with {\it ad hoc} probability distributions. 
The LF Hamiltonian formulation of quantum field theory provides a natural formalism to compute 
hadronization at the amplitude level.~\cite{Brodsky:2008tk}  In this case, one uses light-front time-ordered perturbation theory for the QCD light-front Hamiltonian to generate the off-shell  quark and gluon T-matrix helicity amplitude  using the LF generalization of the Lippmann-Schwinger formalism:
\begin{equation}
T ^{LF}= 
{H^{LF}_I }  \\ + 
{H^{LF}_I }{1 \over {\cal M}^2_{\rm Initial} - {\cal M}^2_{\rm intermediate} + i \epsilon} {H^{LF}_I }  
+ \cdots 
\end{equation}
Here   ${\cal M}^2_{\rm intermediate} \!  = \! \sum^N_{i=1} {(\mbf{k}^2_{\perp i} + m^2_i )/x_i}$ is the invariant mass squared of the intermediate state and ${H^{LF}_I }$ is the set of interactions of the QCD LF Hamiltonian in the ghost-free light-cone gauge.~\cite{Brodsky:1997de}
The $T^{LF}$ matrix element is
evaluated between the out and in eigenstates of $H^{QCD}_{LF}$.   The event amplitude generator is illustrated for $e^+ e^- \to \gamma^* \to X$ in Fig. \ref{hadroniz}.

The LFWFS of AdS/QCD can be used as the interpolating amplitudes between the off-shell quark and gluons and the bound-state hadrons.
Specifically,
if at any stage a set of  color-singlet partons has  light-front kinetic energy 
$\sum_i {\mbf{k}^2_{\perp i}/ x_i} \!  < \! \Lambda^2_{\rm QCD}$, then one coalesces the virtual partons into a hadron state using the AdS/QCD LFWFs.   This provides a specific scheme for determining the factorization scale which  matches perturbative and nonperturbative physics.

This scheme has a number of  important computational advantages:

(a) Since propagation in LF Hamiltonian theory only proceeds as $\tau$ increases, all particles  propagate as forward-moving partons with $k^+_i \ge 0$.  There are thus relatively few contributing
 $\tau$-ordered diagrams.

(b) The computer implementation can be highly efficient: an amplitude of order $g^n$ for a given process only needs to be computed once.  In fact, each non-interacting cluster within $T^{LF}$ has a numerator which is process independent; only the LF denominators depend on the context of the process.  This method has recently been used by   L.~Motyka and A.~M.~Stasto~\cite{Motyka:2009gi}
to compute gluonic scattering amplitudes in QCD.

(c) Each amplitude can be renormalized using the ``alternate denominator'' counterterm method, rendering all amplitudes UV finite.~\cite{Brodsky:1973kb}

(d) The renormalization scale in a given renormalization scheme  can be determined for each skeleton graph even if there are multiple physical scales.

(e) The $T^{LF}$ matrix computation allows for the effects of initial and final state interactions of the active and spectator partons. This allows for leading-twist phenomena such as diffractive DIS, the Sivers spin asymmetry and the breakdown of the PQCD Lam-Tung relation in Drell-Yan processes.

(f)  ERBL and DGLAP evolution are naturally incorporated, including the quenching of  DGLAP evolution  at large $x_i$ where the partons are far off-shell.

(g) Color confinement can be incorporated at every stage by limiting the maximum wavelength of the propagating quark and gluons.

(h) This method retains the quantum mechanical information in hadronic production amplitudes which underlie Bose-Einstein correlations and other aspects of the spin-statistics theorem.
Thus Einstein-Podolsky-Rosen QM correlations are maintained even between far-separated hadrons and  clusters.

A similar off-shell T-matrix approach was used to predict antihydrogen formation from virtual positron--antiproton states produced in $\bar p A$ 
collisions.~\cite{Munger:1993kq}

\section{Dynamical Effects of Rescattering \label{rescat}}

Initial-state and
final-state rescattering, neglected in the parton model, have a profound effect in QCD hard-scattering reactions,
predicting single-spin asymmetries,~\cite{Brodsky:2002cx,Collins:2002kn} diffractive deep lepton-hadron inelastic scattering,~\cite{Brodsky:2002ue} the breakdown of
the Lam Tung relation in Drell-Yan reactions,~\cite{Boer:2002ju} nor nuclear shadowing and non-universal 
antishadowing~\cite{Brodsky:2004qa}---leading-twist physics which is not incorporated in
the light-front wavefunctions of the target computed in isolation. 
It is thus important to distinguish~\cite{Brodsky:2008xe} ``static'' or ``stationary'' structure functions which are computed directly from the LFWFs of the target  from the ``dynamic'' empirical structure functions which take into account rescattering of the struck quark.   Since they derive from the LF eigenfunctions of the target hadron, the static structure functions have a probabilistic interpretation.  The wavefunction of a stable eigenstate is real; thus the static structure functions cannot describe diffractive deep inelastic scattering nor the single-spin asymmetries since such phenomena involves the complex phase structure of the $\gamma^* p $ amplitude.  
One can augment the light-front wavefunctions with a gauge link corresponding to an external field
created by the virtual photon $q \bar q$ pair
current,~\cite{Belitsky:2002sm,Collins:2004nx} but such a gauge link is
process dependent,~\cite{Collins:2002kn} so the resulting augmented
wavefunctions are not universal.~\cite{Brodsky:2002ue,Belitsky:2002sm,Collins:2003fm}

It should be emphasized
that the shadowing of nuclear structure functions is due to the
destructive interference between multi-nucleon amplitudes involving
diffractive DIS and on-shell intermediate states with a complex
phase.  The physics of rescattering and shadowing is thus not
included in the nuclear light-front wavefunctions, and a
probabilistic interpretation of the nuclear DIS cross section is
precluded. 
In addition, one finds that antishadowing in deep inelastic lepton-nucleus scattering is is not universal,~\cite{Brodsky:2004qa}
but depends on the flavor of each quark and antiquark struck by the lepton.  Evidence of this phenomena has been reported by 
Scheinbein {\it et al}.~\cite{Schienbein:2009kk}

\begin{figure}[!]
\centering
\includegraphics[width=13cm]{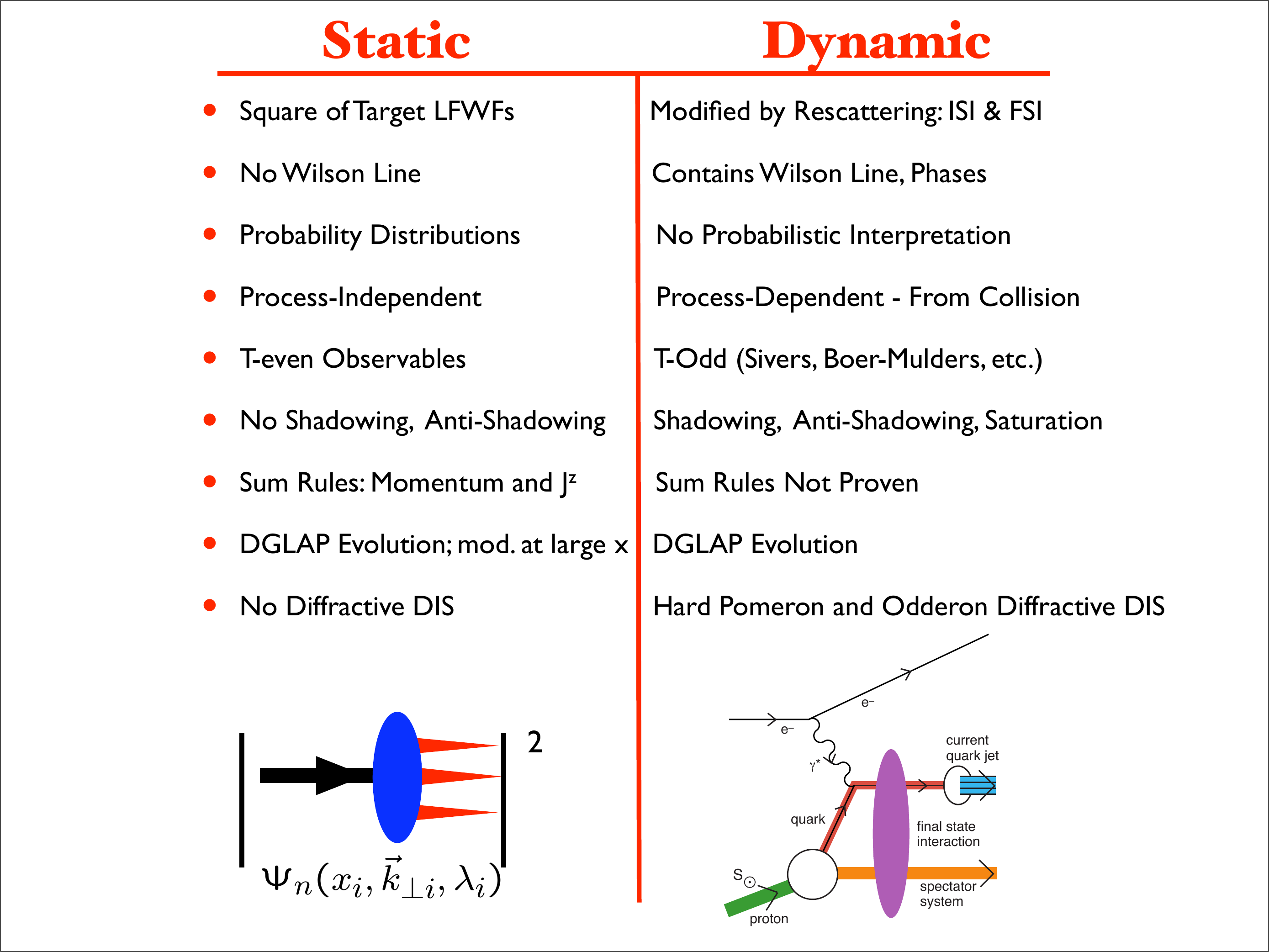}
\caption{Static vs dynamic structure functions}
\label{figstatdyn}  
\end{figure}

The distinction 
between static structure functions; i.e., the probability distributions  computed from the square of the light-front wavefunctions, versus the nonuniversal dynamic structure functions measured in deep inelastic scattering is summarized in Fig. \ref{figstatdyn}.

\section{Novel Perspectives on QCD from Light-Front Dynamics}

In this section we summarize a number of  topics  where new, and in some cases surprising, perspectives for QCD physics have emerged from the light-front formalism.

\begin{enumerate}

\item  It is natural to assume that the nuclear modifications to the structure functions measured in deep inelastic lepton-nucleus and neutrino-nucleus interactions are identical;  in fact,  the Gribov-Glauber theory predicts that the antishadowing of nuclear structure functions is not  universal, but depends on the quantum numbers of each struck quark and antiquark.~\cite{Brodsky:2004qa}  This observation can explain the recent analysis of Schienbein {\it et al.},~\cite{Schienbein:2008ay} which shows that the NuTeV measurements of nuclear structure functions obtained from neutrino  charged current reactions differ significantly from the distributions measured in deep inelastic electron and muon scattering.

\item It is conventional to assume that high transverse momentum hadrons in inclusive high energy hadronic collisions,  such as $ p p \to H X$,  only arise  from jet fragmentation. In fact, a significant fraction of high $p^H_\perp$ events  can emerge  directly from the hard subprocess
itself.~\cite{Arleo:2009ch,Arleo:2010yg}  This phenomena can explain~\cite{Brodsky:2008qp} the  ``baryon anomaly'' observed at RHIC-- the ratio of baryons to mesons at high $p^H_\perp$,  as well as the power-law fall-off $1/ p_\perp^n$ at fixed $x_\perp = 2 p_\perp/\sqrt s, $ increases with centrality,~\cite{Adler:2003kg} opposite to the usual expectation that protons should suffer more energy loss in the nuclear medium than mesons.

\item The effects of final-state interactions of the scattered quark  in deep inelastic scattering  have been traditionally assumed to be power-law suppressed.  In fact,  the final-state gluonic interactions of the scattered quark lead to a  $T$-odd non-zero spin correlation of the plane of the lepton-quark scattering plane with the polarization of the target proton.~\cite{Brodsky:2002cx}  This  leading-twist ``Sivers effect''  is non-universal since QCD predicts an opposite-sign correlation~\cite{Collins:2002kn,Brodsky:2002rv} in Drell-Yan reactions, due to the initial-state interactions of the annihilating antiquark. 
The final-state interactions of the struck quark with the spectators~\cite{Brodsky:2002ue}  also lead to diffractive events in deep inelastic scattering (DDIS) at leading twist,  such as $\ell p \to \ell^\prime p^\prime X ,$ where the proton remains intact and isolated in rapidity;    in fact, approximately 10 \% of the deep inelastic lepton-proton scattering events observed at HERA are
diffractive.~\cite{Adloff:1997sc, Breitweg:1998gc} The presence of a rapidity gap
between the target and diffractive system requires that the target
remnant emerges in a color-singlet state; this is made possible in
any gauge by the soft rescattering incorporated in the Wilson line or by augmented light-front wavefunctions.  

\item It is usually assumed --  following the intuition of the parton model -- that the structure functions  measured in deep inelastic scattering can be computed in the Bjorken-scaling leading-twist limit from the absolute square of the light-front wavefunctions, summed over all Fock states.  In fact,  dynamical effects, such as the Sivers spin correlation and diffractive deep inelastic lepton scattering due to final-state gluon interactions,  contribute to the experimentally observed DIS cross sections. 
Diffractive events also lead to the interference of two-step and one-step processes in nuclei which in turn, via the Gribov-Glauber theory, lead to the shadowing and the antishadowing of the deep inelastic nuclear structure functions;~\cite{Brodsky:2004qa}  such phenomena are not included in the light-front wavefunctions of the nuclear eigenstate.
This leads to an important  distinction between ``dynamical''  vs. ``static''  (wavefunction-specific) structure functions.~\cite{Brodsky:2009dv}

\item
As  noted by Collins and Qiu,~\cite{Collins:2007nk} the traditional factorization formalism of perturbative QCD  fails in detail for many hard inclusive reactions because of initial- and final-state interactions.  For example, if both the
quark and antiquark in the Drell-Yan subprocess
$q \bar q \to  \mu^+ \mu^-$ interact with the spectators of the
other  hadron, then one predicts a $\cos 2\phi \sin^2 \theta$ planar correlation in unpolarized Drell-Yan
reactions.~\cite{Boer:2002ju}  This ``double Boer-Mulders effect" can account for the large $\cos 2 \phi$ correlation and the corresponding violation~\cite{Boer:2002ju, Boer:1999mm} of the Lam Tung relation for Drell-Yan processes observed by the NA10 collaboration.   
An important signal for factorization breakdown at the LHC  will be the observation of a $\cos 2 \phi$ planar correlation in dijet production.

\item	  It is conventional to assume that the charm and bottom quarks in the proton structure functions  only arise from gluon splitting $g \to Q \bar Q.$  In fact, the proton light-front wavefunction contains {\it ab initio } intrinsic heavy quark Fock state components such as $\vert uud c \bar c\rangle$~\cite{Brodsky:1980pb,Brodsky:1984nx,Harris:1995jx,Franz:2000ee}.   The intrinsic heavy quarks carry most of the proton's momentum since this minimizes the off-shellness of the state. The heavy quark pair $Q \bar Q$ in the intrinsic Fock state  is primarily a color-octet,  and the ratio of intrinsic charm to intrinsic bottom scales scales as $m_c^2/m_b^2 \simeq 1/10,$ as can easily be seen from the operator product expansion in non-Abelian QCD.   Intrinsic charm and bottom explain the origin of high $x_F$ open-charm and open-bottom hadron production, as well as the single and double $J/\psi$ hadroproduction cross sections observed at high $x_F$.   The factorization-breaking nuclear $A^\alpha(x_F)$ dependence  of hadronic $J/\psi$ production cross sections is also explained.  A novel mechanism for inclusive and diffractive
Higgs production $pp \to p p H  $, in which the Higgs boson carries a significant fraction of the projectile proton momentum, is discussed in Ref.~\cite{Brodsky:2006wb}  The production
mechanism is based on the subprocess $(Q \bar Q) g \to H $ where the $Q \bar Q$ in the $\vert uud Q \bar Q \rangle$ intrinsic heavy quark Fock state of the colliding proton has approximately
$80\%$ of the projectile protons momentum. 

\item	It is often stated that the renormalization scale of the QCD running coupling $\alpha_s(\mu^2_R) $  cannot be fixed, and thus it has to be chosen in an {\it ad hoc} fashion.  In fact, as in QED, the scale can be fixed unambiguously by shifting $\mu_R$  so that all terms associated with the QCD $\beta$ function vanish.  In general, each set of skeleton diagrams has its respective scale. The result is independent of the choice of the initial renormalization scale ${\mu_R}_0$, thus satisfying Callan-Symanzik invariance.  Unlike heuristic scale-setting procedures,  the BLM method~\cite{Brodsky:1982gc} gives results which are independent of the choice of renormalization scheme, as required by the transitivity property of the renormalization group.   The divergent renormalon terms of order $\alpha_s^n \beta^n n!$ are transferred to the physics of the running coupling.  Furthermore, one retains sensitivity to ``conformal''  effects which arise in higher orders; physical effects which are not associated with QCD  renormalization.  The BLM method also provides scale-fixed,
scheme-independent high precision connections between observables, such as the ``Generalized Crewther Relation'',~\cite{Brodsky:1995tb} as well as other ``Commensurate Scale Relations''.~\cite{Brodsky:1994eh,Brodsky:2000cr}  Clearly the elimination of the renormalization scale ambiguity would greatly improve the precision of QCD predictions and increase the sensitivity of searches for  new physics at the LHC.

\item It is usually assumed that the QCD coupling $\alpha_s(Q^2)$ diverges at $Q^2=0$; {\it i.e.},``infrared slavery''.  In fact, determinations from lattice gauge theory,  Bethe-Salpeter methods, effective charge measurements, gluon mass phenomena, and AdS/QCD all lead (in their respective scheme) to a finite value of the QCD coupling in the infrared.~\cite{Brodsky:2010ur}  Because of color confinement, the quark and gluon propagators vanish at long 
wavelength: $k < \Lambda_{QCD}$~\cite{bjorken}, and consequently the quantum loop corrections underlying the  QCD $\beta$-function  decouple in the infrared, and  the coupling  freezes to a finite value at 
$Q^2 \to 0$.~\cite{Brodsky:2007hb,Brodsky:2008be}   This observation underlies the use of conformal methods in AdS/QCD.

\item It is conventionally assumed that the vacuum of QCD contains quark $\langle 0 \vert q \bar q \vert 0 \rangle$ and gluon  $\langle 0 \vert  G^{\mu \nu} G_{\mu \nu} \vert 0 \rangle$ vacuum condensates, although the resulting vacuum energy density leads to a $10^{45}$  order-of-magnitude discrepancy with the 
measured cosmological constant.~\cite{Brodsky:2009zd}  However, a new perspective has emerged from Bethe-Salpeter and light-front analyses where the QCD condensates are identified as ``in-hadron'' condensates, rather than  vacuum entities, but consistent with the Gell Mann-Oakes-Renner  relation.~\cite{Brodsky:2010xf} The ``in-hadron''  condensates become realized as higher Fock states of the hadron when the theory is quantized at fixed light-front time $\tau = x^0 + x^3.$

\item  In nuclear physics nuclei are composites of nucleons. However, QCD provides a new perspective:~\cite{Brodsky:1976rz,Matveev:1977xt}  six quarks in the fundamental
$3_C$ representation of $SU(3)$ color can combine into five different color-singlet combinations, only one of which corresponds to a proton and
neutron.  The deuteron wavefunction is a proton-neutron bound state at large distances, but as the quark separation becomes smaller,
QCD evolution due to gluon exchange introduces four other ``hidden color'' states into the deuteron
wavefunction.~\cite{Brodsky:1983vf} The normalization of the deuteron form factor observed at large $Q^2$,~\cite{Arnold:1975dd} as well as the
presence of two mass scales in the scaling behavior of the reduced deuteron form factor,~\cite{Brodsky:1976rz} suggest sizable hidden-color
Fock state contributions  in the deuteron
wavefunction.~\cite{Farrar:1991qi}
The hidden-color states of the deuteron can be materialized at the hadron level as   $\Delta^{++}(uuu)\,  \Delta^{-}(ddd)$ and other novel quantum
fluctuations of the deuteron. These dual hadronic components become important as one probes the deuteron at short distances, such
as in exclusive reactions at large momentum transfer.  For example, the ratio  ${{d \sigma/ dt}(\gamma d \to \Delta^{++}
\Delta^{-})/{d\sigma/dt}(\gamma d\to n p) }$ is predicted to increase to  a fixed ratio $2:5$ with increasing transverse momentum $p_T.$
Similarly, the Coulomb dissociation of the deuteron into various exclusive channels $e d \to e^\prime + p n, p p \pi^-, \Delta \, \Delta, \cdots$
will have a changing composition as the final-state hadrons are probed at high transverse momentum, reflecting the onset of hidden-color
degrees of freedom.

\item It is usually assumed that the imaginary part of the deeply virtual Compton scattering amplitude is determined at leading twist by  generalized parton distributions, but that the real part has an undetermined  ``$D$-term'' subtraction. In fact, the real part is determined by the  local  two-photon interactions of the quark current in the QCD light-front Hamiltonian.~\cite{Brodsky:2008qu,Brodsky:1971zh}  This contact interaction leads to a real energy-independent contribution to the DVCS amplitude  which is independent of the photon virtuality at fixed  $t$.  The interference of the timelike DVCS amplitude with the Bethe-Heitler amplitude leads to a charge asymmetry in $\gamma p \to \ell^+ \ell^- p$.~\cite{Brodsky:1971zh,Brodsky:1973hm,Brodsky:1972vv}   Such measurements can verify that quarks carry the fundamental electromagnetic current within hadrons.

\end{enumerate}

\section{Conclusions}
The fifth dimension of anti-de Sitter space provides a remarkable mathematical tool for studying the behavior of hadrons in physical space and time as the length scale changes. Light-front holography makes this correspondence specific -- the variable $z$ in AdS space becomes uniquely identified with the Lorentz-invariant variable  $\zeta$, the coordinate which measures the separation of the quark and gluonic constituents within a hadron at equal light-front time $\tau$. The mapping of
the expressions for computing electromagnetic and gravitational form factors in AdS space to the
corresponding expressions in light-front Hamiltonian theory confirms this
correspondence.

The identification of the coordinate $z$ in AdS space  with $\zeta$ at fixed light-front time also provides a physical understanding of the dynamics described by AdS/QCD. The  $\zeta$ dependence of the relativistic light-front wave equations determines the off-shell dynamics of the bound states as a function of the invariant mass of the constituents.  The variable $L$, which appears as a parameter in the five-dimensional mass parameter  $\mu R$ in AdS space, is identified as the kinematic orbital angular momentum $L^z$ of the constituents in 3+1 space at fixed light-front time.

A long-sought goal in hadron physics is to find a simple analytic first approximation to QCD analogous to the Schr\"odinger-Coulomb equation of atomic physics.	This problem is particularly challenging since the formalism must be relativistic, color-confining, and consistent with chiral symmetry.
We have shown that
the AdS wave equations, modified by a  non-conformal dilaton background field  which incorporates the confinement interaction, leads, via light-front holography, to a simple 
Schr\"odinger-like light-front wave equation.
The result is a single-variable
light-front wave equation in $\zeta$ with an effective confining potential, which determines the eigenspectrum and the light-front wavefunctions of hadrons for general spin and orbital angular momentum. 
In fact, $\zeta$ plays the same role in relativistic quantum field theory as the radial coordinate $r$ of non-relativistic Schr\"odinger quantum mechanics.   

For a positive-dilaton profile, a remarkable one-parameter description of nonperturbative hadron dynamics is obtained.~\cite{deTeramond:2008ht, deTeramond:2005su, Brodsky:2010px} This  model predicts a zero-mass pion for zero-mass quarks and a Regge spectrum of linear trajectories with the same slope in the (leading) orbital angular momentum $L$ of the hadrons and their radial  quantum number $N$.
Given the light-front wavefunctions, one can compute  a wide range of hadron properties, including decay constants, structure functions, distribution amplitudes and hadronic form factors.   The AdS/QCD light-front wavefunctions also lead to a method for computing the hadronization of quark and gluon jets at the amplitude level.~\cite{Brodsky:2008tk}

Light-front AdS/QCD implements chiral symmetry  in a novel way:~\cite{Brodsky:2010px}
the hadron eigenstates generally have components with different orbital angular momentum; e.g.,  the proton eigenstate in AdS space with massless quarks has $L=0$ and $L=1$ light-front Fock components with equal probability.    The effects of chiral symmetry breaking increase as one goes toward large interquark separation, consistent with spectroscopic data.

The AdS/QCD soft-wall model also predicts the form of the non-perturbative effective coupling $\alpha_s^{AdS}(Q)$ and its $\beta$-function.~\cite{Brodsky:2010ur}
The AdS/QCD model can be systematically improved  by using its complete orthonormal solutions to diagonalize the full QCD light-front Hamiltonian~\cite{Vary:2009gt} or by applying the Lippmann-Schwinger method in order to systematically include the QCD interaction terms. 

We have also reviewed some novel features  of QCD,  including
the consequences of confinement for quark and gluon condensates. 
The distinction
between static structure functions, such as the probability
distributions  computed from the square of the light-front
wavefunctions, versus dynamical structure functions which include the
effects of rescattering, has also emphasized. We have also discussed the relevance of the light-front Hamiltonian formulation of QCD to describe the
coalescence of quark and gluons into hadrons.~\cite{Brodsky:2008tk}

\section*{Acknowledgments}
Presented by SJB at Light Cone 2010: Relativistic Hadronic and Particle Physics,  June 14-18, 2010,
Valencia, Spain.    We are grateful to the organizers of LC2010 for  their outstanding hospitality, and we thank our collaborators for many helpful conversations. 
This research was 
supported by the Department of Energy  contract DE--AC02--76SF00515.  SLAC-PUB-14275.

\end{document}